\documentclass[preprint, superscriptaddress]{revtex4-1}
\usepackage{float}
\usepackage{longtable}
\usepackage{mathtools}
\usepackage[utf8]{inputenc}
\usepackage[T1]{fontenc}
\usepackage{lmodern}
\usepackage{url}
\usepackage{color}
\usepackage{setspace}

\begin{document}

\title{Defect Tolerant Monolayer Transition Metal Dichalcogenides}
\author{Mohnish Pandey}
\affiliation{Center for Atomic-scale Materials Design, Department of
Physics, Technical University of Denmark, DK - 2800 Kongens Lyngby,
Denmark}
\author{Filip A. Rasmussen}
\affiliation{Center for Atomic-scale Materials Design, Department of
Physics, Technical University of Denmark, DK - 2800 Kongens Lyngby,
Denmark}
\author{Korina Kuhar}
\affiliation{Center for Atomic-scale Materials Design, Department of
Physics, Technical University of Denmark, DK - 2800 Kongens Lyngby,
Denmark}
\author{Thomas Olsen}
\affiliation{Center for Atomic-scale Materials Design, Department of
Physics, Technical University of Denmark, DK - 2800 Kongens Lyngby,
Denmark}
\author{Karsten W. Jacobsen}
\affiliation{Center for Atomic-scale Materials Design, Department of
Physics, Technical University of Denmark, DK - 2800 Kongens Lyngby,
Denmark}
\author{Kristian S. Thygesen}
\affiliation{Center for Atomic-scale Materials Design, Department of
Physics, Technical University of Denmark, DK - 2800 Kongens Lyngby,
Denmark}
\affiliation{Center for Nanostructured Graphene (CNG),
Department of Physics, Technical University of Denmark, DK - 2800 Kongens Lyngby, Denmark}

\email{thygesen@fysik.dtu.dk}
\date{\today}

\begin{abstract}
Localized electronic states formed inside the band gap of a semiconductor
due to crystal defects can be detrimental to the material’s opto-electronic
properties. Semiconductors with lower tendency to form defect induced deep gap states are termed
defect tolerant. Here we provide a systematic first-principles investigation of defect
tolerance in 29 monolayer transition metal dichalcogenides
(TMDs) of interest for nano-scale optoelectronics. We find that the TMDs based on group VI and X metals form deep
gap states upon creation of a chalcogen (S, Se, Te) vacancy while the TMDs based on group IV metals form only
shallow defect levels and are thus predicted to be defect tolerant. Interestingly, all the defect sensitive
TMDs have valence and conduction bands with very similar orbital composition. This indicates a
bonding/anti-bonding nature of the gap which in turn suggests that dangling bonds will fall inside the gap.
These ideas are made quantitative by introducing a descriptor that measures the degree of similarity of the
conduction and valence band manifolds. Finally, the study is generalized to non-polar nanoribbons of the TMDs
where we find that only the defect sensitive materials form edge states within the band gap. 

\end{abstract}
\maketitle

Single layers of semi-conducting transition metal dichalcogenides (TMDs)
are presently attracting much attention due to their unique opto-electronic properties which include
direct-indirect band gap transitions\cite{mak_atomically_2010, splendiani_emerging_2010, zeng_optical_2013,
zhang_direct_2014-1}, valley selective spin-orbit interactions,\cite{cao_valley-selective_2012,
mak_control_2012} high charge carrier mobilities,\cite{kaasbjerg_phonon-limited_2012,
kaasbjerg_acoustic_2013, Multi-Terminal-Transport-Measurements_2015} and strong light-matter interactions
arising from large oscillator strengths and tightly bound excitons.\cite{britnell_strong_2013, bernardi_extraordinary_2013}
Another attractive feature of the two-dimensional (2D) materials is that their electronic
properties can be readily tuned e.g. by applying strain
\cite{conley_bandgap_2013}, electrostatic gating \cite{liu_tuning_2012,
rostami_effective_2013}, or by van der Waals heterostructure engineering\cite{gao_artificially_2012,
geim_van_2013,DielectricGenomeHeterostructure-Nanolett2015}. In parallel with this development new 2D materials are
continuously being discovered. For example, monolayers and multilayers
of MoTe$_2$, NbSe$_2$, NiTe$_2$, TaS$_2$, TaSe$_2$, TiS$_2$, WS$_2$,
WSe$_2$, ZrS$_2$ have recently been synthesized or isolated by exfoliation.\cite{ChemistryDichalco-Chhowalla-NatChem2013}.   

One of the main performance limiting factors of opto-electronics devices such as photo detectors, light emitting diodes,
solar cells, and field effect transistors, is 
the presence of crystal defects in the active semiconductor material. Such defects can act as local scattering centers
which reduce the mobility of
charge carriers and enhance non-radiative recombination of photo-excited electron-hole pairs.
The effectiveness of a defect to scatter charge carriers, trap excitons and induce
recombination between electrons and holes depends crucially on the way the defect
modifies the electronic structure around the band edges; in particular whether or
not it introduces localized states inside the band gap (deep gap states).
On the other hand, defect states lying close, i.e. on the order of $k_B T$, to the conduction or
valence band edges (shallow defect states) might enhance the charge carrier concentration and thereby
the conductivity.\cite{C5TC03109E} Depending on the extent of localization of a defect state it might
affect the mobility as well. The materials where defects introduce deep gap states are termed defect
sensitive, while the materials where no or only shallow levels appear are called defect tolerant.

There have been several experimental \cite{DefectDominatedDopingMoS2-ACSNano2014,
HoppingTransportLocalizedStatesMoS2-NatComm2013} and theoretical
\cite{NativeDdefectsBulkMonolayerMoS2-PRB2015} studies of defects and their
influence on the electronic properties of few-layer MoS$_2$ - the most well
studied of the TMDs. These studies indicate that S vacancies are the most common
type of defects and that they lead to the formation of localized states inside
the band gap. The sulphur vacancies are believed to be the main reason for
the low mobility observed in back gated field effect transistors based on chemical
vapour deposition (CVD) grown MoS$_2$, which is usually 1-2 orders of
magnitude lower than the theoretical limit set by phonon scattering
\cite{kaasbjerg_phonon-limited_2012, kaasbjerg_acoustic_2013}. In contrast, 
mobilities very close to the theoretical limit were recently measured in
van der Waals heterostructure devices where a high quality mechanically exfoliated MoS$_2$ monolayer was
encapsulated into hexagonal boron nitride and contacted by graphene electrodes
\cite{Multi-Terminal-Transport-Measurements_2015}.
Defect-induced deep gap states are also responsible for the ultrafast non-radiative recombination of
photo-excited excitons in MoS$_2$ which limits the quantum efficiency of CVD grown TMDs to <0.01.\cite{doi:10.1021/nl503636c, UltrasensitivePhotodetectorsMonolayerMoS2,
ElectricallyTunableWSe2PNJunctions}
Indeed, it was recently demonstrated that chemical passivation of the dangling bonds around the S vacancies in MoS$_2$ increases
the photoluminescence quantum efficiency to almost unity.\cite{doi:10.1021/acs.nanolett.5b00952, Amani1065} 
However, defects and impurity doping can also be used constructively. For example, p-type conductivity
in MoS$_2$, which is otherwise naturally n-doped,
has been recently explored via Niobium doping which introduces shallow acceptor
levels near the valence band edge. \cite{doi:10.1021/nl503251h,
C4RA12498G}. Defect states in monolayer WSe2,
\cite{Voltage-controlledQuantumLightAtomicallyThinSemiconductor}
and hBN \cite{QuantumEmissionHexagonalBNMonolayers} have recently been shown
to act as single photon emitters with exciting opportunities for quantum technology.
Engineering of the chemical activity of monolayer MoS$_2$ for
the hydrogen evolution reaction was recently demonstrated via tuning the
concentration of sulfur vacancies. \cite{ActivatingMoS2BasalPlanes-NatMater2015}  

In this Letter, we systematically explore the tolerance of monolayer TMDs
to chalcogen vacancies. Using ab-initio methods we calculate band structures
of 29 monolayer semiconducting TMDs with and without chalcogen
vacancies. The compounds have been selected from a 2D materials database which
contains different electronic properties calculated from first-principles DFT and GW methods.
\cite{Computational2DMaterialsDatabase-JPCC2015}
The correlation between the tendency of TMDs to form deep gap states and the orbital
character of the valence and conduction band is established via
a simple descriptor, the normalized orbital overlap (NOO), which is calculated from the projected
density of states of the pristine system. 
We find that deep gap states are introduced for all the TMDs based on group VI and group X metals
valence and conduction bands have similar orbital character (NOO close to 1)
while no states or only shallow states are introduced for the other materials
(NOO significantly less than 1). Additionally,
we explore nanoribbons of all the TMDs and find that cleaving of the monolayer
along a non-polar direction of the defect tolerant TMDs induces only
shallow states in the band gap as opposed to the nanoribbons of defect
sensitive TMDs which all have metallic edge states.

\begin{figure}[H]
  \begin{center}
    \includegraphics[width=!, height=6.5cm]{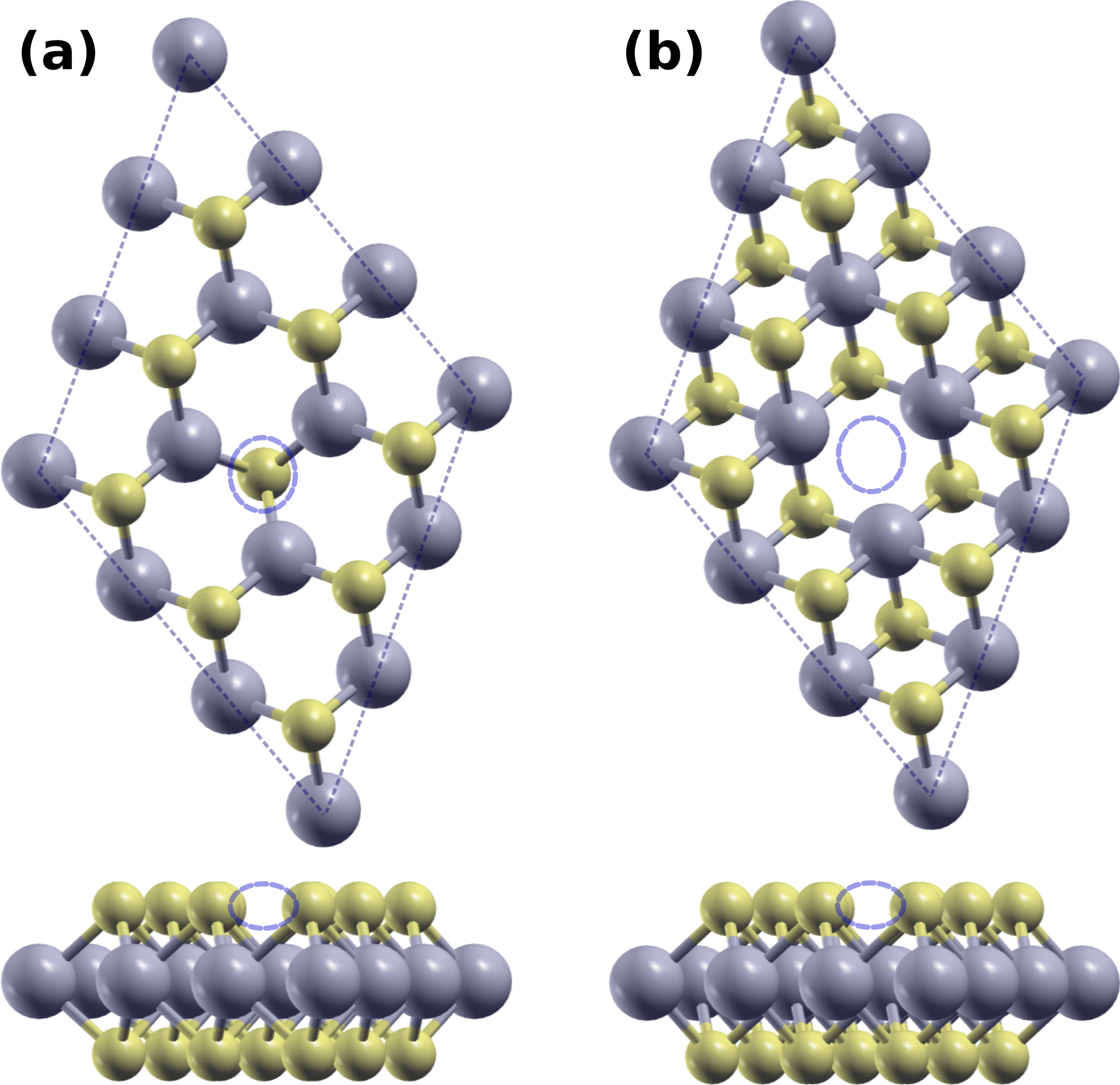}
  \end{center}
  \caption{(a) Top and sideview of a prototypical structure of a metal
dichalcogenide in the 2H structure
used for the defect calculations. The unit cell is shown by the dotted lines.
The defect structure is created by making a
chalcogen vacancy as indicated by a blue ellipse. (b) Similar figure as (a) for the
1T structure.}
  \label{structure}
\end{figure}

All density functional theory (DFT) calculations were performed with the GPAW electronic structure code
\cite{enkovaara_electronic_2010}. The wave functions are expanded
on a real space grid with a grid spacing of 0.18 $\AA$ and we use the
PBE exchange-correlation (xc)-functional \cite{perdew_generalized_1996}. All the pristine structures
have been relaxed until the forces on each atom were below 0.05 eV/$\AA$. Structures with chalcogen
vacancies were modelled using a $3\times 3$ supercell, see Figure \ref{structure}.

In this work we base our analysis of defect states on the PBE single-particle band structures.
This may seem as a drastic oversimplification as the PBE is known to underestimate the band gap of
semiconductors, and its description of localised states is also problematic due to self-interaction errors
which tend to push occupied levels up in energy. However, due to the large number of systems investigated
here, and because we are interested in the qualitative features of the band structure, i.e. whether
or not the vacancies introduce localized states in
the gap, rather than the absolute energies of band edges and gap states, we rely on the PBE band structures in
the present work. As discussed below we find that the PBE results are qualitatively fully consistent with the more rigorous Slater-Janak theory.

The standard way to analyse defect energy levels in semi-conductors is based on total energy
calculations for supercells containing the defect in different charge states. This procedure is, however, not
straightforward. For example, it does not overcome the PBE band gap problem, and thus the correct band
edge positions must be inferred from experiments or more accurate calculations such as the GW method.
\cite{huser_how_2013}
Moreover, the slow convergence of the total energy of a charged supercell with the cell size implies that
some kind of energy correction scheme must be applied to achieve meaningful numbers and there is no unique
solution to this problem. \cite{0022-3719-18-5-005, PhysRevB.51.4014, PhysRevB.78.235104,PhysRevLett.102.016402,
PhysRevB.86.045112, PhysRevLett.84.1942} Alternatively, Slater-Janak (SJ) transition state theory may be used to
obtain the defect levels without the need to compare total energies of differently charged systems.\cite{Pacchioni,
JanakJCP2012} The SJ method exploits that the Kohn-Sham eigenvalues are related to the derivative of the total energy
$E$ with respect to the occupation number $\eta_i$ of the respective orbital,
\begin{equation}\label{eq:SJ}
\frac{\partial E[N]}{\partial \eta_i} = \varepsilon_i.
\end{equation}
Assuming that the eigenvalue for the highest occupied state $\varepsilon_H$ varies linearly with the occupation
number (this is in fact a very good approximation), the electron affinity level can be obtained as 
\begin{equation}
E^{N+1}-E^N = \int_0^1 \varepsilon_H(\eta) d\eta = \varepsilon_H(\frac{1}{2}).
\end{equation}
Thus one obtains the electron affinity level (ionization potential) as the highest unoccupied (lowest occupied)
single-particle eigenvalue of the system with 0.5 electron added to (removed from) the supercell. The SJ model
has previously been shown to predict semiconductor defect levels in good agreement
with the results obtained from more elaborate total energy difference schemes.\cite{LiNbO3}

For all 29 TMDs, we have applied the SJ method to compute the ionization potential and electron affinity levels
of a $3\times 3$ supercell containing a chalcogen vacancy. For consistency the systems have been structurally
relaxed with the added $\pm 0.5$ electron in the supercell. The results do not differ qualitatively from those
derived from the PBE band structure of the neutral supercells. In particular, the two methods predict the same
set of materials to be defect tolerant and defect sensitive, respectively. However, there are small differences between the two approaches.
Taking MoS$_2$ as an example, the neutral PBE spectrum shows an occupied defect level positioned around 0.2
eV above the top of the valence band, see Fig. 2(a). After removal of 0.5 electron, the self-consistent PBE spectrum no
longer shows this defect level. This indicates that the PBE description
places the occupied defect level too high in energy which is most likely due to the PBE self-interaction error. Interestingly,
PBE total energy difference calculations find the defect state below the valence band maximum in agreement
with the SJ calculation.\cite{NativeDdefectsBulkMonolayerMoS2-PRB2015} However, this disagreement with the neutral PBE spectrum does
not affect the conclusions regarding the defect tolerance of MoS$_2$ because both PBE and SJ consistently
predict the presence of unoccupied defect levels at around 0.7 eV below the conduction band minimum (see Fig. 2(a))
which again agrees with calculations based on total energy differences.\cite{NativeDdefectsBulkMonolayerMoS2-PRB2015}

We mention that PBE+U calculations were also performed for all the TMDs with U values ranging from 0 to 4 eV. However,
the band gaps of the group VI and group X TMDs were found to decrease with increasing U leading to unphysical
small band gaps for common U values. Based on these findings, we do not consider PBE+U to be more accurate than
PBE for the considered class of TMDs. 
 
\begin{figure}[H]
  \begin{center}
    \includegraphics[width=12cm, height=!]{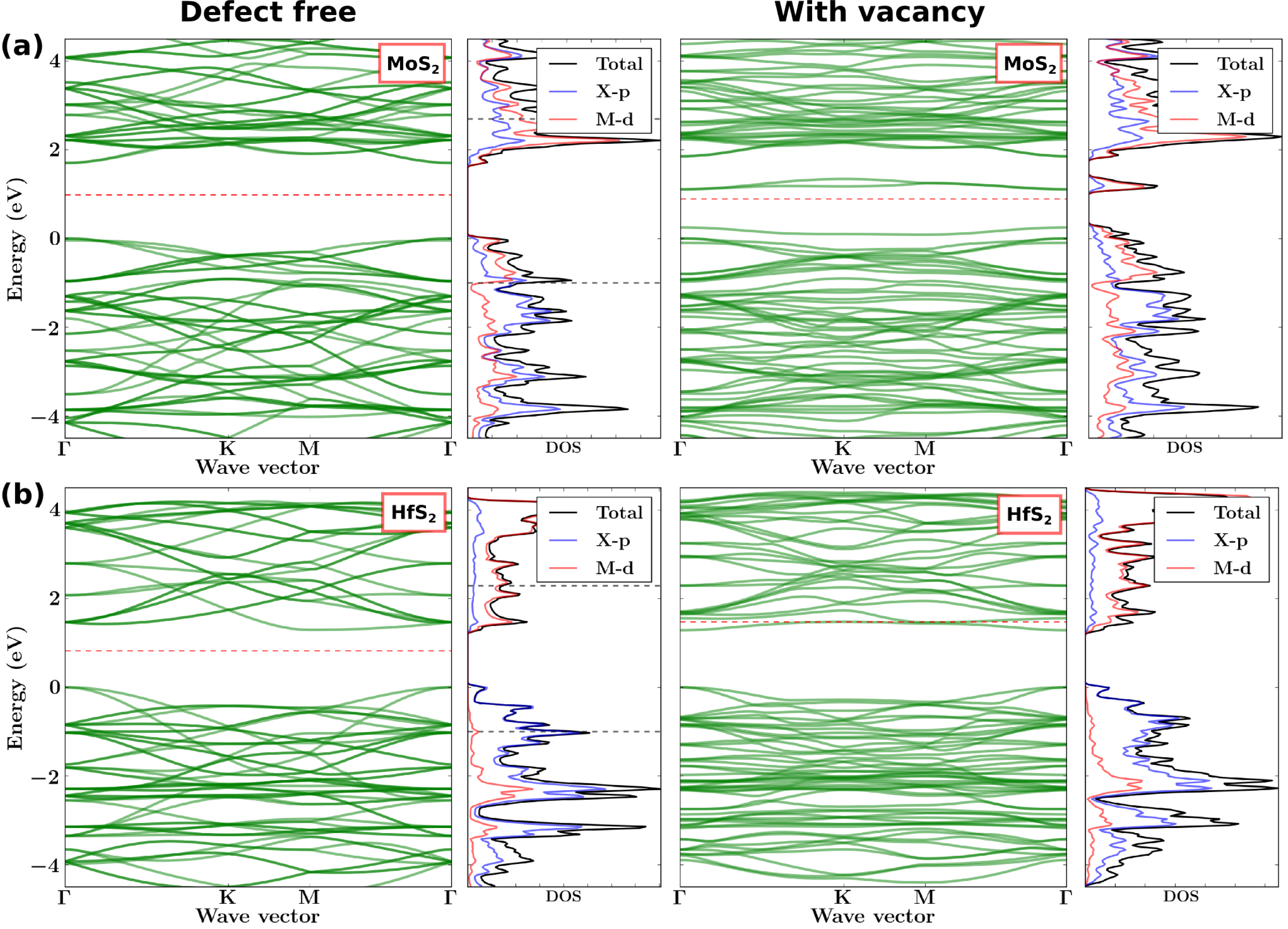}
  \end{center}
  \caption{(a)PBE band structures, the DOS and the
PDOS  on the chalcogen
$p$-orbitals and metal $d$-orbitals of the monolayer
2H-MoS$_2$ in its pristine form
(left) and and with an S vacancy (right). For comparison,
the band structure is plotted for 3$\times$3 supercell for the pristine as
well as the defect structure.
The energy levels have been
aligned to the valence band maximum of the pristine monolayer. (b) Similar figure
as (a) for 1T-HfS$_2$}
  \label{2H-Mo-1T-Hf}
\end{figure}

Figure \ref{2H-Mo-1T-Hf} (a) shows the PBE band structures,
the total density of states (DOS) and the projected density of states
(PDOS) on the chalcogen $p$-orbitals and metal $d$-orbitals
of 2H-MoS$_2$ in its pristine form
(left) and with an S vacancy (right).
The three narrow bands inside the band gap (one just above the valence band edge and two almost degenerate
bands just above the band gap center) are formed by dangling Mo $d$ bonds localized around the S vacancy. From the PDOS it is seen that the
valence band maximum (VBM) and conduction band minimum (CBM) have very similar
orbital character indicating that they consist of
bonding and anti-bonding combinations of sulphur $p$ and metal $d$ states,
respectively. We note that the finite size of the supercell (3$\times$3) is
responsible for the small dispersion of the deep gap states. We have found that
applying a 4$\times$4 supercell does not change the conclusions.

Figure \ref{2H-Mo-1T-Hf} (b) shows a similar plot as Figure \ref{2H-Mo-1T-Hf} (a)
for 1T-HfS$_2$. In contrast to MoS$_2$ this compound largely
conserves its electronic structure around the band edges and no defect state is introduced. 
Additionally, the DOS plot shows that the states near the VBM are mostly dominated by the
chalcogen $p$ states whereas the CBM edge states mainly consist of the metal $d$ states.

\begin{figure}[H]
  \begin{center}
    \includegraphics[width=8.2cm, height=!]{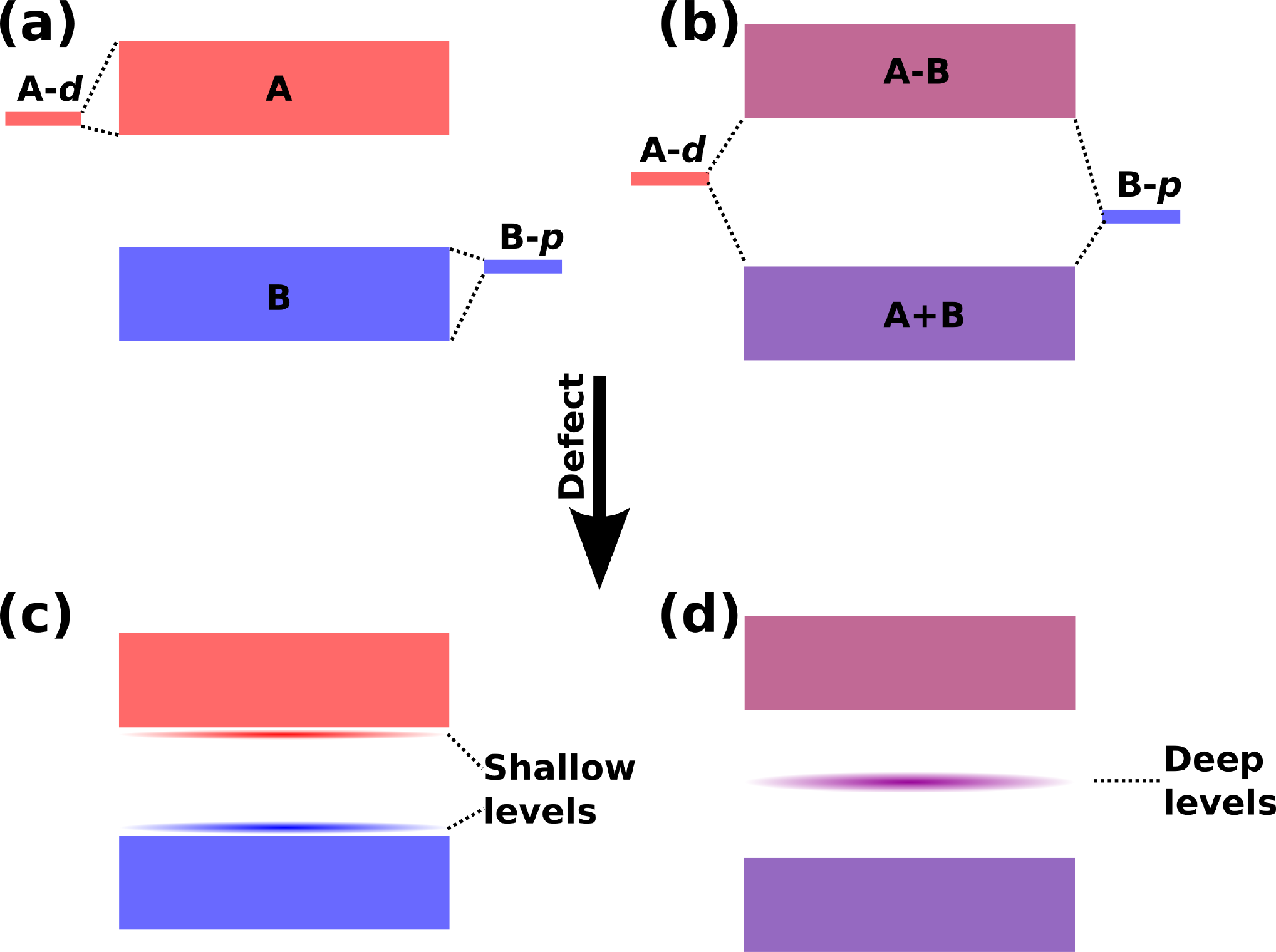}
  \end{center}
  \caption{(a) and (b) show the nature of the band structures near the band edges for
the defect tolerant and defect sensitive cases respectively. In all the TMDs studied here,
the states near the band edges primarily have contributions from the metal $d$
states and chalcogen $p$ states. In the defect tolerant case the nature of the bands
near the band edges are significantly different whereas in the defect sensitive case they
are of mixed nature. (c) and (d) show the shallow and deep levels introduced after the
creation of defects.}
  \label{sketch}
\end{figure}

The above examples indicate that the orbital character of the valence/conduction bands
are crucial for the tendency to form deep gap states. Based on the above picture the
materials can be categorized in two classes; one with conduction and valence
bands composed of bonding and anti-bonding combinations of similar orbitals, the other with valence
and conduction bands composed of orbitals with distinctly different character, see
Figure \ref{sketch}.
 This picture also agrees with previous works on defect tolerance
in semiconductors. \cite{DefectTolerantSemiconductorsSolar-JPCL2014}
The correspondence between the orbital character of the bands and the defect tolerance can be understood
from elementary chemical bond theory. In the case of similar orbital character of
the valence and conduction bands, the former will have bonding character and the
latter will have the anti-bonding character. This implies that the energy of the individual orbitals
should lie inside the band gap, see Figure \ref{sketch} (b). Thus the creation of a vacancy
leaves a dangling bond state corresponding to one of the orbitals lying deep in the
band gap. On the other hand, different orbital character of the valence and conduction bands suggests that these bands are
formed by orbitals lying outside the band gap, and therefore
dangling bond states should not fall inside the gap.
To quantify the orbital character
of the electronic states in a given energy window from $E_1$ to E$_2$ we introduce the
orbital fingerprint vector,
\begin{align}
    |\alpha\rangle &= \frac{1}{\sqrt{c}}\begin{bmatrix}
           \rho_{\nu_1} \\
           \rho_{\nu_2} \\
           \vdots \\
           \rho_{\nu_N}
         \end{bmatrix}
\label{alpha}
\end{align}
where $c$ is a normalization constant, the $\nu_i$'s are (a, l) pairs where `a' and `l' denote
the atom and angular momentum channel, respectively;
and $\rho_{\nu_i}$ is the projected
density of states onto the atomic orbital
$\phi_{\nu_i}$ integrated over the energy window,
\begin{eqnarray}
\rho_{\nu_i} = \sum_n \int_{E_1}^{E_2} |\langle \psi_n | \phi_{\nu_i }\rangle|^2 \delta(E-E_n) dE  
\label{rho}
\end{eqnarray}

Using the orbital fingerprint vector we can define the normalized orbital
overlap (NOO) between two manifolds
of bands located in the energy windows $E^v_1$ to $E^v_2$ and $E^c_1$ to $E_2^c$ as
\begin{equation}
D = \langle\alpha|\beta\rangle
\label{distance}
\end{equation}
where $\alpha$ and $\beta$ correspond to the valence and conduction band manifolds.
By taking the two energy windows to lie around the VBM and CBM, respectively,
we have a measure of the difference in average orbital character around the
valence and conduction band edges. We note that $D=0$ for materials with completely
different character of the valence and conduction bands while $D=1$ for materials
with identical orbital character at the valence and conduction bands.

\begin{figure}[H]
  \begin{center}
    \includegraphics[width=12cm, height=!]{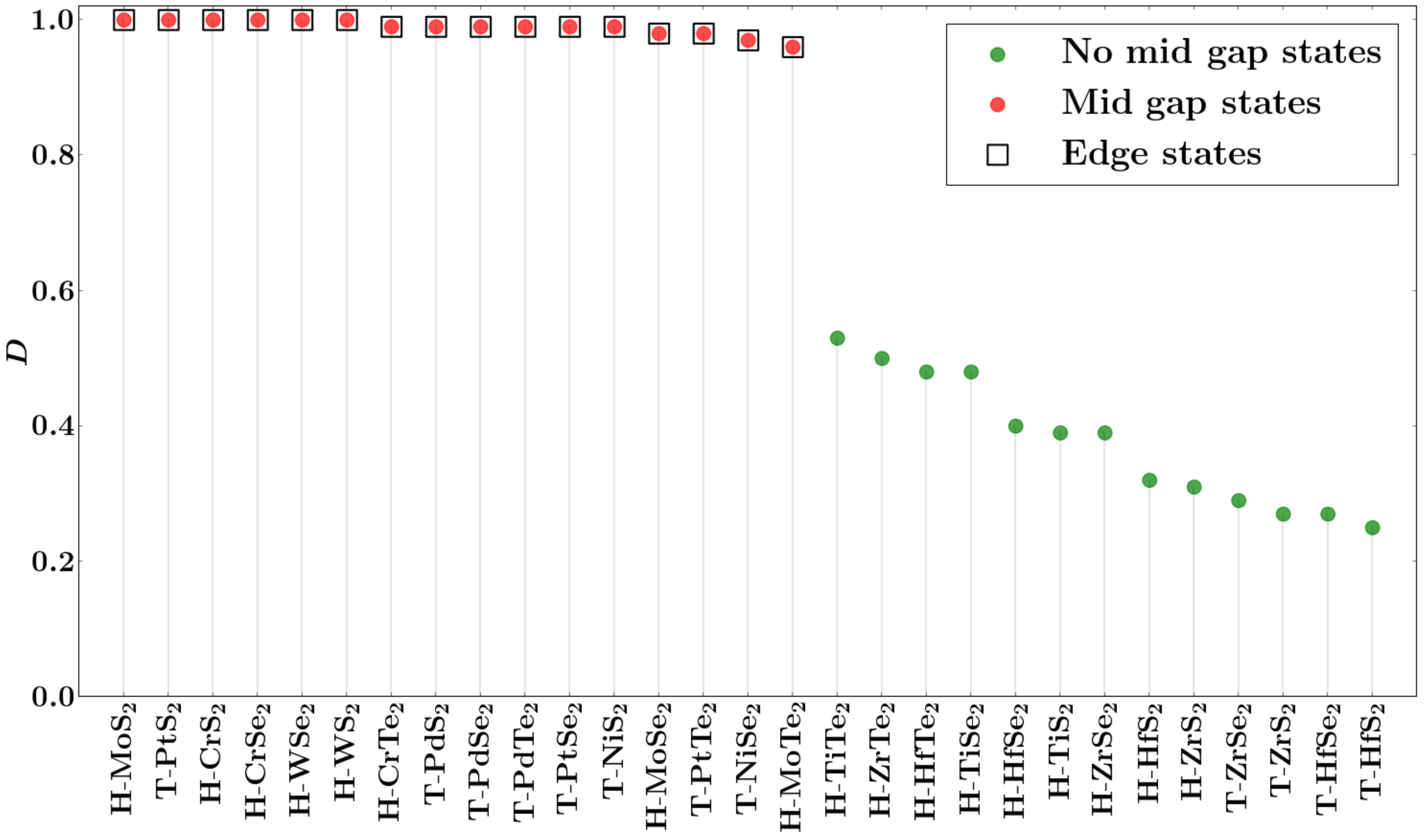}
  \end{center}
  \caption{The plot showing the compounds (y-axis) and the corresponding
$D$ values (x-axis). Red circles indicate the compounds manifesting deep defect
states and the green circles indicating the compounds showing no deep defect states
after the removal of a chalcogen atom. The black squares represent compounds showing states
appearing deep in the band gap after cleaving the monolayer to form nanoribbons.
Cr, Mo and W dichalcogenides are group-IV, Ti, Zr
and Hf dichalcogenides are group-VI and Ni, Pd and Pt dichalcogenides are group-X.}
  \label{descriptor}
\end{figure}

We have computed $D$ (with an energy window of 1 eV above/below the
conduction/valence band extrema) for a set
of 29 monolayer TMDs, see Figure \ref{descriptor}.
We have carried out band structure calculations
like the ones shown in Figure \ref{2H-Mo-1T-Hf} for all 29 TMDs.
For all the materials with $D\sim 1$ we find
localized states inside the band gap. In contrast  all the materials with $D$ significantly
less than 1 do not show deep gap states.
The compounds exhibiting deep defect states
after the removal of a chalcogen atom are shown with red circles in the
Figure \ref{descriptor} whereas the compounds where no deep defect states
are introduced after a chalcogen atom is removed are marked with green
circles. This clearly shows that the normalized orbital overlap
between the valence and conduction band manifolds represent a reliable and
quantitative descriptor for the degree of defect tolerance of the TMDs.
We stress that this rule should apply to the case of vacancies, crystal
distortions, or other perturbations whose effect is to distort the intrinsic
bonding. On the other hand, in the case of impurity atoms, the presence of
deep gap states depends also on the energy of the atomic orbitals of the
impurity atom relative to the band edges.

\begin{figure}[H]
  \begin{center}
    \includegraphics[width=8.2cm, height=!]{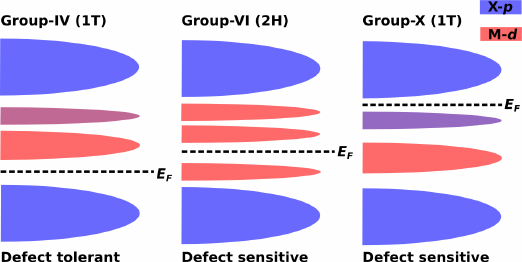}
  \end{center}
  \caption{The sketch showing the filling of the electronic levels of the different groups
of the transition metal dichalcogenides. The TMDs of group-IV are defect tolerant whereas
the TMDs of group-VI $\&$ X are defect sensitive. The chalcogen $p$ states are shown
in blue and the metal $d$ states in red. The regions with mixed color have mixed $p$ and $d$
character.}
  \label{grouping}
\end{figure}

A clear division of 29 TMDs into defect tolerant and defect sensitive
materials can be understood from Figure \ref{grouping}. The splitting
of the levels, the relative contribution of the chalcogen-$p$ and metal-$d$ 
and the position of the Fermi level decides the value of the descriptor. For
example, in the 1T the position of the Fermi level in group-4 TMDs
leads to the VBM having chalcogen-$p$ and the CBM having metal-$d$ character
hence leading to defect tolerance. On the other hand, filling up of more
levels in group-10 compounds places the Fermi level in such a way that the
VBM and CBM states have dominant chalcogen-$p$ character (for example, see
SI for PDOS of NiS$_2$) resulting in similar orbital fingerprint vectors
consequently leading to defect sensitivity. A similar argument can be applied
to the compounds of group-6 in the 2H structure.

Additionally, we also calculated the band structure
of nanoribbons cleaved from the monolayer TMDs. As recently explained, TMD
nanoribbons cleaved along a polar direction will manifest metallic edge states due
to the presence of a dipole across the ribbon. \cite{MetallicStatesMoS2-Marzari-Nanolett2015}
However, cleaving the monolayer along a non-polar direction can introduce
edge states due to the formation of dangling bonds thus having a close
resemblance to the case of a monolayer with a vacancy where the shallow/deep
levels arise due to the presence of dangling bonds. Therefore, we expect
that the arguments for the monolayers with vacancies will also be applicable
for the edges in the nanoribbons. 

\begin{figure}[H]
  \begin{center}
    \includegraphics[width=8.2cm, height=!]{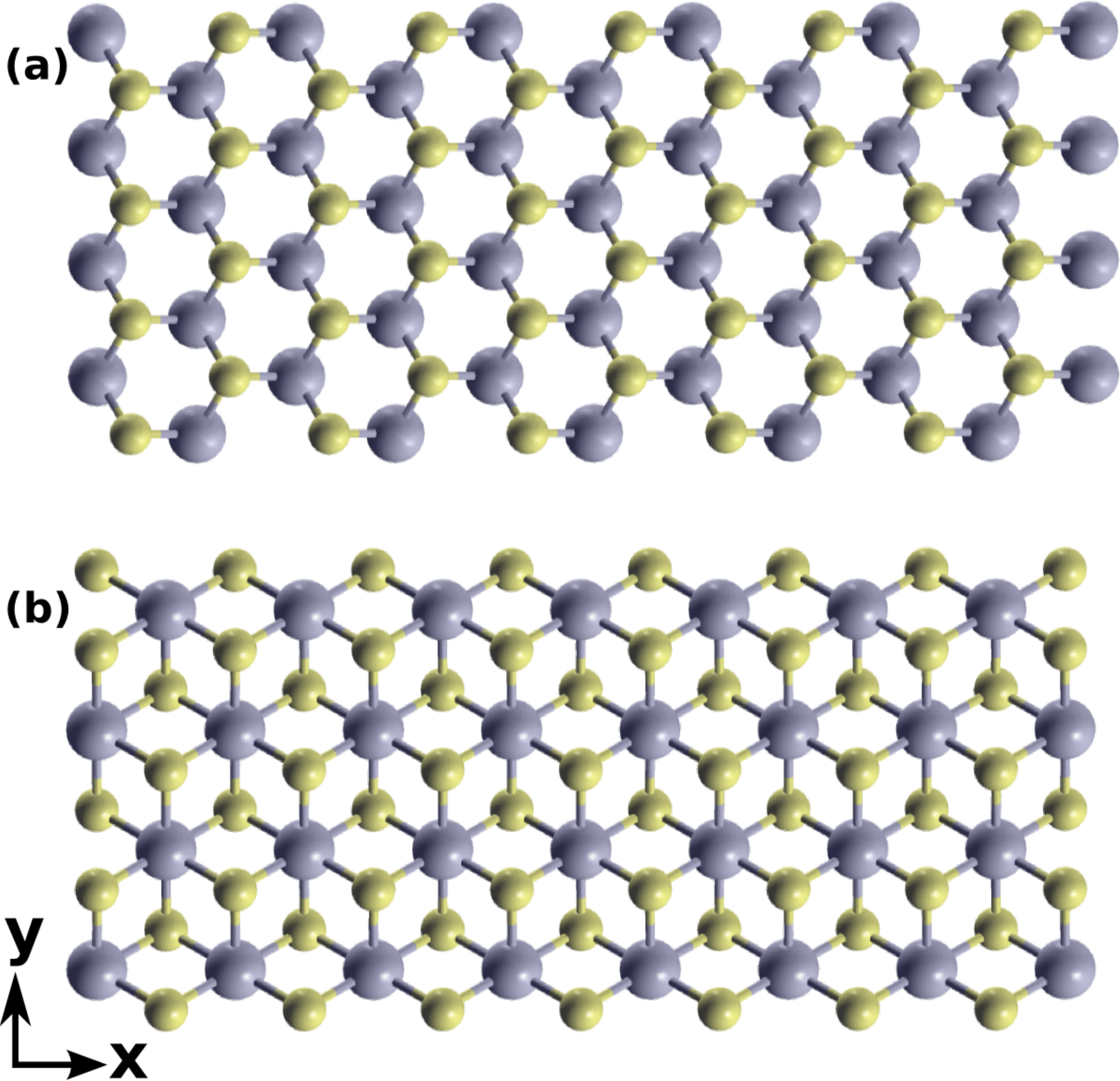}
  \end{center}
  \caption{(a) shows the structure of a nanoribbon of the 2H structure
cleaved along the non-polar direction. The edges of the nanoribbons lie
along the x-axis with a finite width along the y-axis; (b) shows a nanoribbon
cleaved from the 1T structure.}
  \label{ribbon}
\end{figure}

Figure \ref{ribbon} (a) and (b) show the structure of the non-polar nanoribbons of
the 2H and 1T structures, respectively, used for the analysis of the edge states.
The edges of the nanoribbons are directed along the  horizontal axis.   
In Figure \ref{descriptor} the nanoribbons supporting edge states (lying deep) in the band
gap are shown with open squares. The figure consistently shows that only the defect
sensitive TMDs form edge states deep in the band gap whereas the defect
tolerant TMDs form only shallow, or no, edge states in the band gap. This indicates that the analysis based on the NOO
descriptor is rather general and should be applicable to other system with imperfections
involving dangling bonds. We are currently investigating the descriptor and possible
generalizations of it for different classes of semiconductors.

In summary, we have explored the sensitivity of the band structure
of 2D TMDs towards chalcogen vacancies. Our analysis shows that the
tendency of the materials to form localized states within the band gap
strongly depends on the similarity of the orbital character
of the states near the conduction and valence band. 
These ideas were made quantitative by introducing a simple descriptor based on
the projected density of states at the conduction and valence band edges. For a
set of 29 semiconducting TMDs we found a clear correlation between the size of
this descriptor and the presence of deep gap states induced by a chalcogen
vacancy or by the formation of a one-dimensional edge. The idea of identifying
defect tolerant materials based on a quantitative descriptor measuring valence and
conduction band state similarity is completely
general. In particular it does not depend on the dimensionality of the material,
and should be useful both as general concept and as a tool for computational
prediction of defect tolerant compounds. \\ [1ex]

\begin{acknowledgments}
The authors acknowledge support from the Danish Council for Independent
Research's Sapere Aude Program, Grant No. 11-1051390. The Center for
Nanostructured Graphene is sponsored by the Danish National Research
Foundation, Project DNRF58.
\end{acknowledgments}
%
\end{document}